\documentclass[twocolumn,showpacs,amsmath,amssymb]{elsart}

\usepackage{amssymb}
\usepackage[dvips]{graphicx}
\begin{document}
\begin{frontmatter}



\title{Vortex and critical fields in charged Bose liquids and unconventional superconductors}


\author[label1]{A.S.~Alexandrov}
\thanks[label1]{supported by the Leverhulme Trust, United Kingdom}
\address{Department of Physics, Loughborough University, Loughborough LE11 3TU, United Kingdom}

\begin{abstract}
A single vortex  in the charged Bose gas (CBG) has a charged core
and its profile  different from the vortex  in neutral and BCS
superfluids. Lower and upper critical fields of CBG are discussed.
The unusual resistive upper critical field, $H_{c2}(T)$, of many
cuprates and a few other unconventional superconductors is
described as the Bose-Einstein condensation field of preformed
bosons-bipolarons. Its nonlinear temperature dependence follows
from the scaling arguments. Exceeding the Pauli paramagnetic limit
is explained. Controversy in the determination of $H_{c2}(T)$ of
cuprates from kinetic and thermodynamic measurements is addressed
in the framework of the bipolaron theory.
\end{abstract}

\begin{keyword}
vortex \sep critical fields \sep bipolarons \sep cuprates
\PACS 74.20.-z \sep 74.72.-h
\end{keyword}
\end{frontmatter}

\section*{Introduction}
 The seminal work by Bardeen,
Cooper and Schrieffer \cite{bcs} taken further by Eliashberg
\cite{eli} to the intermediate coupling solved one of the major
 problems in Condensed Matter Physics. High-temperature
superconductors present a challenge to the conventional theory.
While the BCS theory provides a qualitatively correct description
of some novel superconductors like magnesium diborade and doped
fullerenes (if the phonon dressing of carriers, i.e. polaron
formation is properly taken into account), cuprates remain a
problem. Here strong antiferromagnetic and charge fluctuations and
the Fr\"{o}hlich and Jahn-Teller electron-phonon interactions have
been identified as an essential piece of physics. In particular,
experimental \cite{mih,cal,tim,guo,shen2,ega,mul2,she2}  evidence
for an exceptionally strong electron-phonon interaction in all
high temperature superconductors is now  overwhelming. Our view,
which we discussed in detail elsewhere \cite{book}  is that the
extension of the BCS theory towards the strong interaction between
electrons and ion vibrations  describes the phenomenon naturally.
The high temperature superconductivity exists in the crossover
region of the electron-phonon interaction strength from the
BCS-like to bipolaronic superconductivity as was predicted before
\cite{ale0}, and explored in greater detail after the discovery
\cite{alemot,workshop,dev,dev2}. The low energy physics in this
 regime is that of a charged Bose gas of small bipolarons, which are
real-space bosons dressed by phonons. They are itinerant
quasiparticles existing in the Bloch states at temperatures below
the characteristic phonon frequency. Here I review  the bipolaron
theory of the vortex state.

\section{Charged vortex}
CBG is an extreme type II superconductor, as shown below. We can
analyse a single vortex in CBG and calculate the critical fields
 by solving a stationary equation for the
macroscopic condensate wave function $\psi _{s}(%
{\bf r})$ \cite{alever},
\begin{eqnarray}
&& \left[\frac{[{\bf \nabla +2}ie{\bf A(r)}]^{2}}{{2}m^{\ast
\ast}} +\mu \right] \psi _{s}({\bf r}) \cr &=&
\frac{4e^{2}}{\epsilon _{0}}\int d{\bf r^{\prime }}\frac{|\psi _{s}({\bf %
r^{\prime }},t)|^{2}-n_{b}}{|{\bf r}-{\bf r^{\prime }|}}
\psi _{s}(%
{\bf r}).
\end{eqnarray}
Subtracting  $n_{b}$ in the integral of Eq.(1)  explicitly takes
into account the Coulomb interaction with the homogeneous charge
background of the same density as the density of charged bosons
$n_{b}$. Here  $2e$ and  $m^{\ast \ast }$ are the charge per boson
and the effective mass, respectively, and $\hbar=c=k_B=1$).

The integra-differential equation (1) is quite different from the
Ginsburg-Landau \cite{gl} and  Gross-Pitaevskii \cite{gp}
equations, describing the vortex in the BCS and netral
superfluids, respectively.  While CBG shares the quantum coherence
with the BCS superconductors and neutral superfluids owing to the
Bose-Einstein condensate (BEC), the long-range (nonlocal)
interaction leads to some peculiarities. In particular, the vortex
is charged in CBG, and the coherence length is just the same as
the screening radius.

Indeed, introducing dimensionless quantities $f=|\psi _{s}|/n_{b}^{1/2}$, $%
{\bf \rho }={\bf r/}\lambda (0)$, and ${\bf h=}2e\xi (0)\lambda
(0)\nabla \times {\bf A}$ for the order parameter, length and
magnetic field, respectively, Eq.(1) and the Maxwell equations
take the following form:
\begin{equation}
{\frac{1}{{\kappa ^{2}\rho }}}{\frac{d}{{d\rho }}}\rho {\frac{df}{{d\rho }}}-%
{\frac{1}{{f^{3}}}}\left( {\frac{dh}{{d\rho }}}\right) ^{2}-\phi
f=0,
\end{equation}
\begin{equation}
{\frac{1}{{\kappa ^{2}\rho }}}{\frac{d}{{d\rho }}}\rho
{\frac{d\phi }{{d\rho }}}=1-f^{2},
\end{equation}
\begin{equation}
{\frac{1}{{\rho }}}{\frac{d}{{d\rho }}}{\frac{\rho }{{f^{2}}}\frac{dh}{{%
d\rho }}}=h.
\end{equation}
A new feature compared with the GL equations for a single vortex
\cite{abr} is the electric field potential determined as
\begin{eqnarray}
\phi &=&{\frac{1}{2e{\phi _{c}}}}\int d{\bf r^{\prime }}V({\bf r}-{\bf %
r^{\prime }})\cr &\times& [|\psi _{s}({\bf r^{\prime
}})|^{2}-n_{b}]
\end{eqnarray}
with a new fundamental unit $\phi _{c}=em^{\ast \ast }\xi
(0)^{2}$. The potential is calculated using the Poisson equation
(3). At $T=0$ the coherence length is the same as the screening
radius,
\begin{equation}
\xi (0)=(2^{1/2}m^{\ast \ast }\omega _{ps})^{-1/2},
\end{equation}
\ and the London penetration depth is
\begin{equation}
\lambda (0)=\left( \frac{m^{\ast \ast }}{16\pi n_{b}e^{2}}\right)
^{1/2}.
\end{equation}
Here $\omega _{ps}=[16 \pi e^2 n_b/(\epsilon_0m^{\ast
\ast})]^{1/2}$ is the plasma frequency. There are now six boundary
conditions in a single-vortex problem. Four of
them are the same as in the BCS superconductor \cite{abr}, $%
h=dh/\rho =0$, $f=1$ for $\rho =\infty $ and the flux quantization
condition, $dh/d\rho =-pf^{2}/\kappa \rho $ for $\rho =0$, where
$p$ is an integer. The remaining two conditions are derived from
the global charge neutrality, $\phi =0$ for $\rho =\infty $ and
\begin{equation}
\phi (0)=\int_{0}^{\infty }\rho \ln (\rho )(1-f^{2})d\rho
\end{equation}
for the electric field at the origin, $\rho =0.$ We notice that
the chemical potential $\mu$ is zero at any point in the thermal
equilibrium.

CBG is an extreme type II superconductor with a very large
Ginsburg-Landau parameter, $\kappa =\lambda (0)/\xi (0)\gg 1$.
Indeed, with the material
parameters typical for oxides, such as $m^{\ast \ast }=10m_{e}$, $%
n_{b}=10^{21}cm^{-3}$ and the static dielectric constant $\epsilon
_{0}=10^{3}$ we obtain $\xi (0)\simeq 0.48$nm, $\lambda (0)\simeq
265$nm, and the Ginsburg-Landau ratio $\kappa \simeq 552$. Owing
to a large dielectric constant the Coulomb repulsion remains weak
even for heavy bipolarons,
\begin{equation}
r_{s}=\frac{4m^{\ast \ast }e^{2}}{\epsilon _{0}(4\pi
n_{b}/3)^{1/3}}\simeq 0.46.
\end{equation}
 If $\kappa \gg 1, $
Eq.(4) is reduced to the London equation with the familiar
solution $h=pK_{0}(\rho )/\kappa $, where $K_{0}(\rho )$ is the
Hankel function of imaginary argument of zero order. For the
region $\rho \leq p$, where the order parameter and the electric
field differ from unity and zero, respectively, we can use the
flux quantization condition to ``integrate out'' the magnetic
field in Eq.(2). That leaves us with two parameter-free equations
written for $r=\kappa \rho $ as
\begin{equation}
\frac{1}{r}\frac{d}{dr}r\frac{df}{dr}-\frac{p^{2}f}{r{^{2}}}-\phi
f=0,
\end{equation}
and
\begin{equation}
\frac{1}{r}\frac{d}{dr}r\frac{d\phi }{dr}=1-f^{2}.
\end{equation}
They are satisfied by regular solutions of the form $f=c_{p}r^{p}$
and $\phi =\phi (0)+(r^{2}/4),$ when $r\rightarrow 0$. The
constants $c_{p}$ and $\phi (0)$ are determined by complete
numerical integration of Eqs.(10) and (11). The numerical results
for $p=1$ are $c_{1}\simeq 1.5188$ and $\phi (0)\simeq -1.0515$.

In the region $p<<r<p\kappa $ the solutions are $f=1+(4p^{2}/r^{4})$ and $%
\phi =-p^{2}/r^{2}$. In this region $f$ differs qualitatively from
the BCS order parameter, $f_{BCS}=1-(p^{2}/r^{2})$ \cite{abr}. The
difference is due to a local charge redistribution caused by the
magnetic field in CBG. Quite different from the BCS
superconductor, where the total density of electrons remains
constant across the sample, CBG allows for flux penetration by
redistributing the density of bosons within the coherence volume.
This leads to an increase of the order parameter compared with the
homogeneous case ($f=1$) in the region close to the vortex core.
Inside the core the order parameter is suppressed, as in the BCS
superconductor. The resulting electric field,  (together with the
magnetic field) acts as an additional centrifugal force increasing
the steepness ($c_{p}$) of the order parameter compared with the
BCS superfluid, where $c_{1}\simeq 1.1664$.

The breakdown of the local charge neutrality is due to the absence
of any equilibrium $normal$ state solution in CBG below
$H_{c2}(T)$ line. Both
superconducting ($\Delta _{{\bf k}}\neq 0$) and normal ($\Delta _{{\bf k}}=0$%
) solutions are allowed at any temperature in the BCS
superconductors. Then the system decides which of two phases (or
their mixture) is energetically favorable, but the local charge
neutrality is respected. In contrast, there is no equilibrium
normal state solution (with $\psi _{s}=0$) in CBG below
$H_{c2}(T)$-line because it does not respect the density sum rule.
Hence, there are no different phases to mix, and the only way to
acquire a flux in the thermal equilibrium is to redistribute the
local density of bosons at the expense of their Coulomb energy.
This energy determines the vortex free energy $ F=E_{v}-E_{0}$,
which is the difference of the energy of CBG with, $E_{v}$, and
without, $E_{0}$, magnetic flux,
\begin{eqnarray}
F&=&\int d{\bf r} {\frac{1}{2m^{\ast }}}|[\nabla +2ie{\bf A(r)}%
]\psi _{s}{\bf (r)}|^{2}\cr
&+& e\phi _{c}\phi \lbrack |\psi _{s}{\bf (r)}%
|^{2}-n_{b}]+{\frac{(\nabla \times {\bf A})^{2}}{{8\pi
}}}\nonumber.
\end{eqnarray}
Using Eqs.(2), (3) and (4) it can be written in the dimensionless
form as
\begin{equation}
F=2\pi \int_{0}^{\infty }[h^{2}-{\frac{1}{{2}}}\phi (1+f^{2})]\rho
d\rho .
\end{equation}
In the large $\kappa $ limit the main contribution comes from the region $%
p/\kappa <\rho <p$, where $f\simeq 1$ and $\phi \simeq
-p^{2}/(\kappa ^{2}\rho ^{2})$. The energy is thus the same as
that in the BCS superconductor, $F\simeq 2\pi p^{2}\ln (\kappa
)/\kappa ^{2}$. It is seen that the most stable solution is the
formation of the vortex with one flux quantum, $p=1,$ and the
lower critical field is the same as in the BCS superconductor,
$h_{c1}\approx \ln \kappa /(2\kappa )$ \cite{abr}. However,
different from the BCS superconductor, where the Ginsburg-Landau
phenomenology is microscopically justified in the temperature
region close to $T_{c}$, the CBG vortex structure is derived here
in the low temperature region. Actually the zero temperature
solution is applied in a wide temperature region well below the
Bose-Einstein condensation temperature, where the depletion of the
condensate remains small. The actual size of the charged core is
about $4\xi $.

\section{Upper critical field in the strong-coupling regime}

If we ``switch off'' the Coulomb repulsion between bosons, an
ideal  CBG cannot be bose-condensed at finite temperatures in a
homogeneous magnetic field because of a one-dimensional particle
motion at the lowest Landau level \cite{scha}. However, an
interacting charged Bose-gas is condensed in a field lower than a
certain critical value $H_{c2}(T)$ \cite{aleH}. Collisions between
bosons and/or with impurities and phonons make the motion
three-dimensional, and eliminate the one-dimensional singularity
of the density of states, which prevents BEC of the ideal gas in
the filed. As we
show below the upper critical field of CBG differs significantly from $%
H_{c2}(T)$ of BCS superconductors. It has an unusual positive
curvature near $T_{c}$, $H_{c2}(T)\sim (T_{c}-T)^{3/2}$ and
diverges at $T\rightarrow 0$, if there is no localisation due to a
random potential. The localization can drastically change the
low-temperature behavior of $H_{c2}(T)$, so that at high density
of impurities a re-entry effect to the normal state might occur.

In line with the conventional definition, $H_{c2}(T)$ is a field,
where a first nonzero solution of the linearized stationary
equation for the macroscopic condensate wave function occurs,
\begin{eqnarray}
&&\left[ {\frac{1}{2m^{\ast \ast }}}[\nabla -2ie{\bf A}({\bf r}%
)]^{2}+\mu \right] \psi _{s}({\bf r})\cr &=&V_{scat}({\bf r}) \psi
_{s}({\bf r}).
\end{eqnarray}
Here we introduce the ``scattering'' potential $V_{scat}({\bf r})$
caused, for example, by particle-particle and/or
particle-impurity collisions.
Let us first discuss noninteracting bosons, $%
V_{scat}({\bf r})=0.$ Their energy spectrum in the homogeneous
magnetic field is
\begin{equation}
\varepsilon _{n}=\omega (n+1/2)+\frac{k_{z}^{2}}{2m^{\ast \ast }},
\end{equation}
where $\omega =2eH_{c2/}m^{\ast \ast }$ and $n=0,1,2,...\infty $.
BEC occurs when the chemical potential ``touches'' the lowest band
edge from below, i.e. $\mu =\omega /2$. Hence, quite different
from the GL equation, the Schr\"{o}dinger equation (13) does not
allow for a direct determination of $H_{c2}$, In fact, it
determines the value of the chemical potential. Then using this
value the upper critical field is found from the density sum rule,
\begin{equation}
\int_{E_{c}}^{\infty }f(\epsilon )N(\epsilon ,H_{c2})d\epsilon
=n_{b},
\end{equation}
where $N(\epsilon ,H_{c2})$ is the density of states (DOS) of the
Hamiltonian, Eq.(13), $f(\epsilon )=[\exp (\epsilon -\mu
)/T-1]^{-1}$ is the Bose-Einstein distribution function, and
$E_{c}$ is the lowest band edge. For ideal bosons we have $\mu
=E_{c}=\omega /2$ and
\begin{eqnarray}
N(\epsilon ,H_{c2})&=&{\frac{\sqrt{2}(m^{\ast \ast })^{3/2}\omega }{{4\pi ^{2}}%
}} \cr &\times& \Re \sum_{n=0}^{\infty }{\frac{1}{\sqrt{\epsilon
-\omega (n+1/2)}}}. \nonumber
\end{eqnarray}
Substituting this DOS into Eq.(15) yields
\begin{eqnarray}
&&{\frac{\sqrt{2}(m^{\ast \ast })^{3/2}\omega }{{4\pi ^{2}}}}\int_{0}^{\infty }%
\frac{dx}{x^{1/2}}\frac{1}{\exp (x/T)-1}\cr
&=&n_{b}-\tilde{n}(T),
\end{eqnarray}
 where
\begin{eqnarray}
\tilde{n}(T)&=&{\frac{\sqrt{2}(m^{\ast \ast })^{3/2}\omega }{{4\pi ^{2}}}}%
\int_{0}^{\infty }\frac{dx}{\exp (x/T)-1} \cr
&\times&
\sum_{n=1}^{\infty }{\frac{1}{\sqrt{%
x-\omega n}}}
\end{eqnarray}
is the number of bosons occupying the levels from $n=1$ to
$n=\infty .$ This
number is practically the same as in zero field, $\tilde{n}%
(T)=n_{b}(T/T_{c})^{3/2}$, if $\omega \ll T_{c}.$ On the contrary,
the number of bosons on the lowest level, $n=0,$ is given by a
divergent integral on the left-hand side of Eq.(18). Hence the
only solution to Eq.(16) is $H_{c2}(T)=0.$

The scattering of bosons effectively removes the one-dimensional
singularity in $N_{0}(\epsilon ,H_{c2})\propto \omega (\epsilon
-\omega /2)^{-1/2}$ leading to a finite DOS near the bottom of the
lowest level,
\begin{equation}
N_{0}(\epsilon ,H_{c2})\propto \frac{H_{c2}}{\sqrt{\Gamma
_{0}(H_{c2})}}.
\end{equation}
Using the Fermi-Dirac golden rule the collision broadening of the
lowest level $\Gamma _{0}(H_{c2})$ is proportional to the same DOS
\begin{equation}
\Gamma _{0}(H_{c2})\propto N_{0}(\epsilon ,H_{c2}),
\end{equation}
so that $\Gamma _{0}$ scales with the field as $\Gamma
_{0}(H_{c2})\propto H_{c2}^{2/3}$. Then the number of bosons at
the lowest level is estimated as
\begin{eqnarray}
n_{0}&=&{\frac{\sqrt{2}(m^{\ast \ast })^{3/2}\omega }{{4\pi ^{2}}}}%
\int_{\Gamma _{0}}^{\infty }\frac{dx}{x^{1/2}}\frac{1}{\exp
(x/T)-1} \cr &\propto& TH_{c2}^{2/3},
\end{eqnarray}
as long as $T\gg \Gamma _{0}.$ Here we apply the one-dimensional
DOS, but cut the integral at $\Gamma _{0}$ from below. Finally we
arrive at
\begin{equation}
H_{c2}(T)=H_{0}(t^{-1}-t^{1/2})^{3/2},
\end{equation}
where $t=T/T_{c}$, and $H_{0}$ is a temperature independent
constant. The
scaling constant $H_{0}$ depends on the scattering mechanism. If we write $%
H_{0}=\Phi _{0}/(2\pi \xi _{0}^{2})$, \ then the characteristic
length is
\begin{equation}
\xi _{0}\approx \left( \frac{l}{n_{b}}\right) ^{1/4},
\end{equation}
where $l$ is the zero-field mean-free path of low energy bosons.

\begin{figure}[tbp]
\begin{center}
\includegraphics[angle=-90,width=0.67\textwidth]{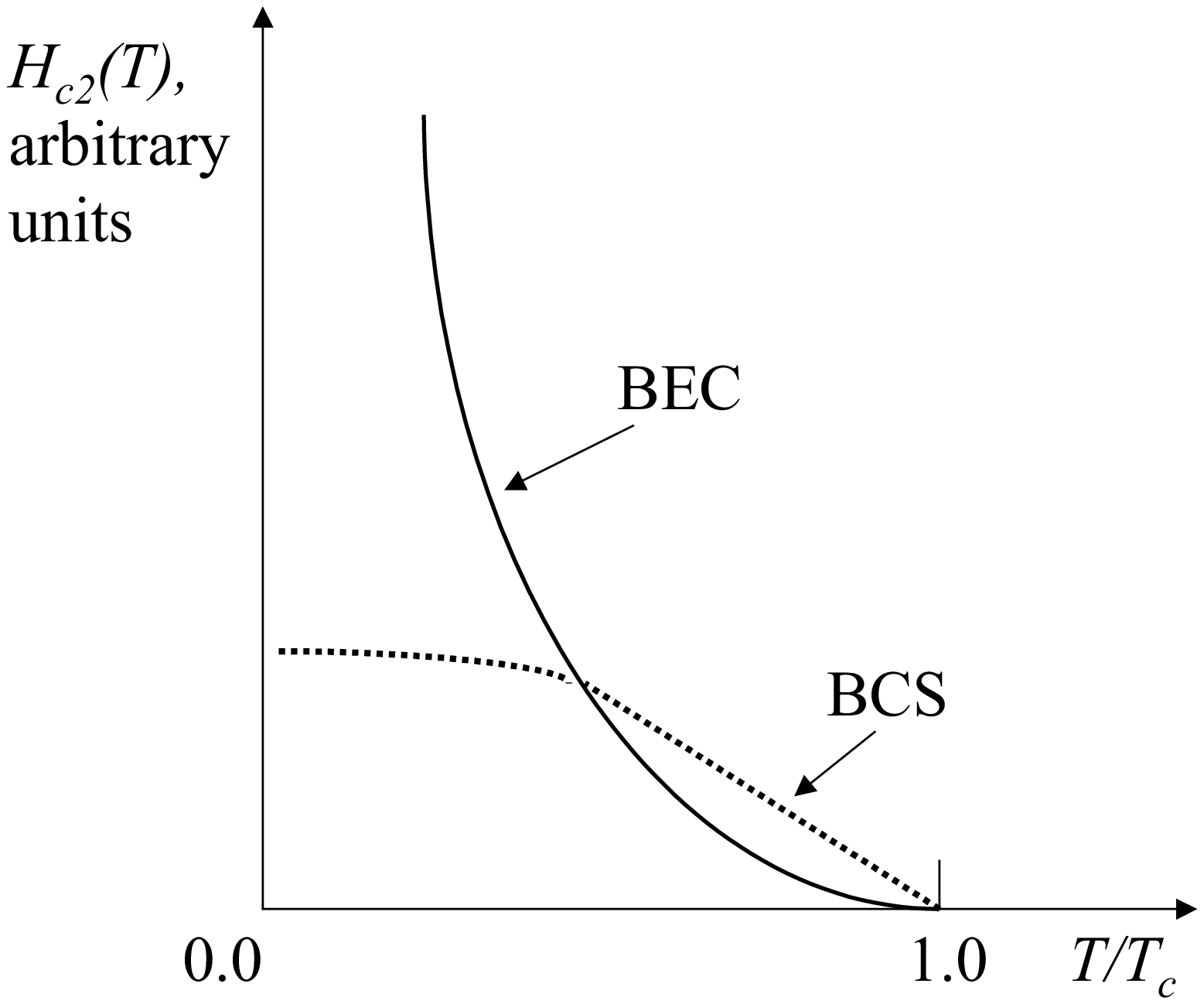} \vskip -0.5mm
\end{center}
\caption{\small {Upper critical field of CBG compared with
$H_{c2}(T)$ of BCS superconductors}.}
\end{figure}
The upper critical field has a nonlinear behaviour,
\[
H_{c2}(T)\propto (T_{c}-T)^{3/2},
\]
in the vicinity of $T_{c}$, and diverges at low-temperatures as
\[
H_{c2}(T)\propto T^{-3/2}.
\]
These simple scaling arguments are fully confirmed by DOS
calculations with impurity \cite{aleH} and boson-boson
\cite{alekabbee} scattering.  The ``coherence'' length $\xi _{0}$
of CBG, Eq.(22), depends on the mean free path $l$ and the
inter-particle distance $n_{b}^{-1/3}.$ It has nothing to do with
the size of the bipolaron, and could be as large as the coherence
length of the weak-coupling BCS superconductors. \

Thus $H_{c2}(T)$ of strongly-coupled superconductors has a ``3/2''
curvature near $T_{c}$ different from the linear BCS $H_{c2}(T)$.
The curvature is a universal feature of CBG, which does not depend
on a particular scattering mechanism and on approximations made.
Another interesting feature of strongly-coupled superconductors is
a breakdown of the Pauli paramagnetic limit given by $H_{p}\simeq
1.84T_{c}$ in the weak-coupling theory. $H_{c2}(T)$ of bipolarons
exceeds this limit because the singlet bipolaron binding energy
$\Delta $ is\ much larger than their $T_{c}.$ Bosons are condensed
at $T=0$ no matter
what their energy spectrum is. Hence, in the charged Bose-gas model, $%
H_{c2}(0)=\infty $, Fig.1. For composed bosons, like bipolarons,
the
pair-breaking limit is given by $\mu _{B}H_{c2}(0)\approx \Delta $, so that $%
H_{c2}(0)\gg H_{p}.$

\section{Universal upper critical field of unconventional superconductors}

In cuprates \cite{gen,mac,oso,alezav,fraH,gan,zavkabale},
spin-ladders \cite {spin} and organic superconductors \cite{org}
high magnetic field studies revealed a non-BCS upward curvature of
resistive $H_{c2}(T)$. When measurements were performed on
low-$T_{c}$ unconventional superconductors
\cite{mac,oso,fraH,spin,org}, the Pauli limit was exceeded by
several times. A non-linear temperature dependence in the vicinity
of $\ T_{c}$ was unambigously observed in a few samples
\cite{alezav,fraH,gan,zavkabale}. Importantly, a thermodynamically
determined $H_{c2}$ turned out much higher than the resistive
$H_{c2}$ \cite{wen} due to contrasting magnetic field dependencies
of the specific heat anomaly and of resistive transition.

We believe that many unconventional superconductors are in the
`bosonic' limit of preformed real-space bipolarons, so their
resistive $H_{c2}$ is actually a critical field of the
Bose-Einstein condensation of charged bosons \cite{aleH}.
Calculations carried out for the heat capacity of CBG (see below)
lead to the conclusion that the resistive $H_{c2}$ and the
thermodynamically determined $H_{c2}$ are very different in
bosonic superconductors. While the magnetic field destroys the
condensate of ideal bosons, it hardly shifts the specific heat
anomaly as observed.

A comprehensive scaling of resistive $H_{c2}$ measurements in
unconventional superconductors is shown in Fig.2 \cite{zavkabale}
in the framework of the microscopic model of charged bosons
scattered off impurities (section 2). Generalised Eq.(21)
accounting for a temperature dependence of the number of
delocalised bosons, $n_{b}(T),$ can be written as
\begin{equation}
H_{c2}(T)=H_{0}\left[
\frac{n_{b}(T)}{tn_{b}(T_{c})}-t^{1/2}\right] ^{3/2}.
\end{equation}
  In the vicinity of $T_{c}$ one obtains the
parameter-free $H_{c2}(T)\propto (1-t)^{3/2}$ using Eq.(23), but
the low-temperature behaviour depends on a particular scattering
mechanism, and a detailed structure of the density of localised
states. As suggested by the normal state Hall measurements in
cuprates  $n_{b}(T)$ can be parameterised as $%
n_{b}(T)=n_{b}(0)+constant\times T$ \cite{alebramot}, so that
$H_{c2}(T)$ is described by a single-parameter expression as
\begin{equation}
H_{c2}(T)=H_{0}\left[ \frac{b(1-t)}{t}+1-t^{1/2}\right] ^{3/2}.
\end{equation}

\begin{figure}[tbp]
\begin{center}
\includegraphics[angle=-0,width=0.57\textwidth]{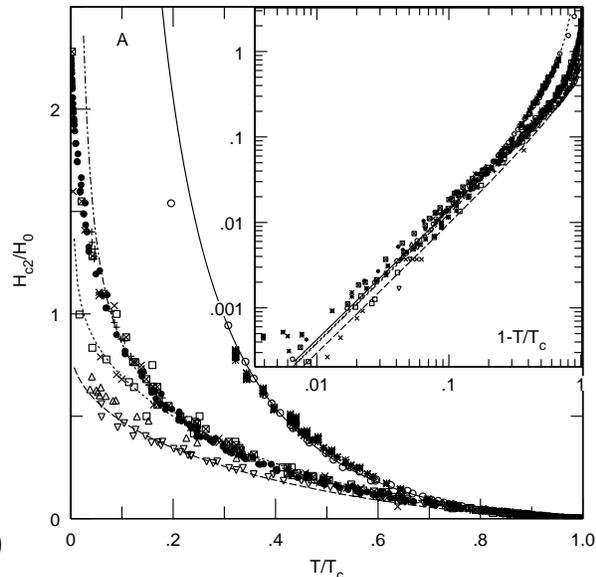} \vskip -0.5mm
\end{center}
\caption{\small{Resisitive upper critical field (determined at
50\% of the transition) of cuprates, spin-ladders and organic
superconductors scaled according to Eq.(24). The parameter b is 1
(solid line), 0.02 (dashed-dotted line), 0.0012 (dotted line), and
0 (dashed line). The inset shows a universal scaling of the same
data near $T_{c}$ on the logarithmic scale. Symbols correspond to
$Tl-2201(\bullet) $, $LSCO (\triangle)$, $Bi-2201 (\times)$,
$Bi-2212 (\ast)$, $YBCO (\circ)$,
$La_{2-x}Ce_{x}CuO_{4-y}(\square)$, $ Sr_2Ca_{12}Cu_{24}O_{41}
(+)$, and $ (TMTSF)_2PF_6 (\triangledown)$ }}
\end{figure}

The parameter $b$ is proportional to the number of delocalised
bosons at zero temperature. We expect that this expression is
applied in the whole temperature region except ultra-low
temperatures, where the Fermi Golden-rule in the scaling fails.
Exceeding the Pauli pair-breaking limit readely follows from the
fact, that the singlet-pair binding energy is related to the
normal-state pseudogap temperature $T^{\ast }$, rather than to
$T_{c}$. $T^{\ast }$ is higher than $T_{c}$ in bosonic
superconductors, and cuprates.

The universal scaling of $H_{c2}$ near $T_{c}$ is confirmed by
resistive measurements of the upper critical field of many
cuprates, spin-ladders, and organic superconductors, as shown in
Fig.2. All measurements reveal a universal $(1-t)^{3/2}$ behaviour
in a wide temperature region (inset), when
they are fitted by Eq.(24). The low-temperature behaviour of $%
H_{c2}(T)/H_{0}$ is not universal, but well described using the
same
equation with the single fitting parameter, $b$. The parameter is close to $%
1 $ in high quality cuprates with a very narrow resistive
transition \cite {gan}. It naturally becomes rather small in
overdoped cuprates where randomness is more essential, so almost
all bosons are localised (at least in one dimension) at zero
temperature.

\section{Specific heat anomaly in CBG}

Bose liquids (or more precisely $He^{4}$) show the characteristic
$\lambda $ -point singularity of their specific heat, but
superfluid Fermi liquids \ like BCS superconductors exhibit a
sharp second order phase transition accompanied by a finite jump
in the specific heat. It was established beyond doubt
\cite{fish,lor,ind,jun,schnel} that the anomaly in high $T_{c}$
cuprates differs qualitatively from the BSC prediction. As was
stressed by Salamon et al.\cite{sal} the heat capacity is
logarithmic near the transition, and consequently, cannot be
adequately treated by the mean-field BCS theory even including the
gaussian fluctuations. In particular, estimates using the gaussian
fluctuations yield an unusually small coherence volume \cite
{lor}, and $Gi$ number of the order of one.

The magnetic field dependence of the anomaly \cite{jun0} is also
unusual, but it can be described by the bipolaron model
\cite{alekablia,zavkabale}. Calculations of the specific heat of
charged bosons in a magnetic field require an analytical DOS,
$N(\epsilon ,B)$ of a particle, scattered by other particles
and/or by a random potential of impurities. We can use DOS in the
magnetic filed with an impurity scattering, which allows for an
analytical result \cite{zavkabale}. The specific heat coefficient
\[
\frac{C(T,B)}{T}=\frac{d}{TdT}\int d\epsilon \frac{N(\epsilon ,B)\epsilon }{%
\exp [(\epsilon -\mu )/T]-1}
\]
calculated with this DOS and with $\mu $ determined from
$n_{b}=\int d\epsilon N(\epsilon ,B)f(\epsilon )$ is shown in
Fig.3.

\begin{figure}[tbp]
\begin{center}
\includegraphics[angle=-0,width=0.57\textwidth]{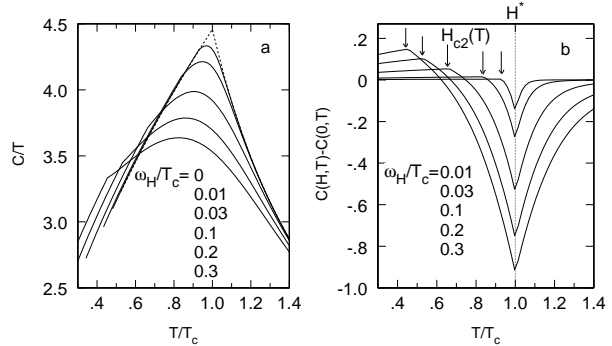} \vskip -0.5mm
\end{center}
\caption{\small {Temperature dependence of the specific heat
devided by temperature (arb. units) of the charged Bose-gas
scattered off impurities for several fields ($\omega
_{_{H}}=2eB/m^{\ast \ast })$. Fig. 3b shows two anomalies, the
lowest one traces resistive transition, while the highest anomaly
is the normal state feature}.}
\end{figure}
The broad maximum at $T\approx T_{c}$ is practically the same as
in the ideal Bose gas without scattering \cite{alekablia}. It
barely shifts in the magnetic field. However, there is the second
anomaly at lower temperatures, which is absent in the ideal gas.
It shifts with the magnetic field, tracing precisely the resistive
transition, as clearly seen from the difference between the
specific heat in the field and zero-field curve, Fig. 3b. The
specific heat, Fig. 3, is in striking resemblance with the Geneva
group's experiments on $DyBa_{2}Cu_{3}0_{7}$ and on
$YBa_{2}Cu_{3}O_{7}$ \cite{jun0}, where both anomalies were
observed.  Within the bipolaron model, when the magnetic field is
applied, it hardly changes the temperature dependence of the
chemical potential near the zero field $T_{c}$ because the energy
spectrum of thermally excited bosons is practically unchanged.
That is because their characteristic energy (of the order of
$T_{c}$) remains huge compared with the magnetic energy of the
order of $2eB/m^{\ast \ast }$. In contrast, the energy spectrum of
low energy bosons is strongly perturbed even by a weak magnetic
field. As a result the chemical potential `touches' the band edge
at lower temperatures, while having almost the same `kink'-like
temperature dependence around $T_{c}$ as in zero field. While the
lower anomaly corresponds to the true long-range order due to the
Bose-Einstein condensation, the higher one is just a `memory'
about the zero-field transition. This microscopic consideration
shows that a genuine phase transition into a superconducting state
is related to resistive transition  and to the lower specific heat
anomaly, while the broad higher anomaly is the normal state
feature of the bosonic system in the external magnetic field.
Different from the BCS superconductor these two anomalies are well
separated in the bosonic superconductor at any field but zero.

In conclusion,  the bipolaron theory of the critical fields and
vortex structures in  strong-coupling superconductors has been
reviewed. A single vortex  in this regime has a charged core and
its profile different from the vortex  in neutral and BCS
superfluids. The upper critical field  is also qualitatively
different from the weak and intermediate-coupling $H_{c2}(T)$. We
have interpreted unusual resistive upper critical fields
 of many  unconventional
superconductors  as the Bose-Einstein condensation field of
preformed bosons-bipolarons. Their nonlinear temperature
dependences follow from the scaling arguments. Exceeding the Pauli
paramagnetic limit has been explained, and the  controversy in the
determination of $H_{c2}(T)$ of cuprates from kinetic and
thermodynamic measurements has been addressed in the framework of
the bipolaron theory.

\end{document}